\documentclass{article}
\usepackage{amssymb}
\usepackage{amsmath}
\usepackage{xcolor}

\begin{document}


\title{The Birth of Quantum Mechanics and the Dirac Equation}

\author{Volodimir Simulik $^{1}$, Denys I. Bondar $^{2*}$\\
$^{1}$Institute of Electron Physics of the NAS of Ukraine\\
21 Universitetska Str. Uzhgorod 88017, Ukraine\\
Email: vsimulik@gmail.com\\
$^{2*}$  Department of Physics and Engineering Physics\\
Tulane University, New Orleans, Louisiana 70118, United States\\
Corresponding author\\
Email: vsimulik@gmail.com}

\date{18. 04. 2026}

\maketitle


\begin{abstract}
The year 2025 marked the centennial of quantum mechanics, inaugurated by
Heisenberg's matrix formulation and the foundational contributions of Pauli,
Schr\"odinger, and Dirac.  Concurrently, 2026 marks the centennial of the
Klein-Gordon equation, the second-order relativistic wave equation from
which both the Schr\"odinger and Dirac equations were derived.  This article
supplements the recent review published in \textit{J.~Phys.\ A: Math.\ Theor.}
\textbf{58} (2025) 053001 by providing a more detailed examination of the
formative period 1925--1928, with particular attention to contributions that
have received insufficient recognition in the standard narrative.  We
reconstruct Kramers' independent derivation of the Dirac equation---obtained
essentially simultaneously with Dirac's own result yet unpublished for seven
years---and discuss its relation to Van der Waerden's group-theoretical
approach.  The role of Charles Galton Darwin in elucidating the physical
content of the Dirac equation is also highlighted.  In addition, we present
two modern derivations not catalogued in the earlier review: one based on
Operational Dynamical Modeling, which deduces the Dirac equation from
relativistic Ehrenfest relations and the canonical commutation algebra, and one
rooted in the Madelung hydrodynamic formulation.  Three broad periods of
quantum theory development---foundational, consolidation, and the modern era
of quantum information---are briefly surveyed.
\end{abstract}



\pagestyle{plain}
\pagenumbering{arabic}
\setcounter{page}{1}

\section{Introduction}
The year 2025 marked the 100th anniversary of the birth of quantum mechanics as a theoretical model of atomic phenomena. A full century has now passed since the publication of Werner Heisenberg's landmark paper~\cite{Real-1} on the matrix formulation of quantum mechanics in terms of observable quantities, together with Wolfgang Pauli's fundamental contributions~\cite{Real-2,Real-3}. Erwin Schr\"odinger's celebrated paper on the fundamental equation of \emph{non-relativistic} quantum mechanics~\cite{Real-4} followed a year later, in 1926. The construction of the foundations of quantum theory was then brought to completion in 1928 with  the publication of Paul Dirac's paper~\cite{Real-5}, which introduced the fundamental equation of \emph{relativistic} quantum mechanics.

Dirac's contribution consists of two parts,~\cite{Real-5} and~\cite{Real-6}. The famous first part~\cite{Real-5} was substantially supplemented by~\cite{Real-6}, in which further problems of the new quantum mechanics were addressed---most notably the Zeeman effect.

It was precisely Heisenberg, Schr\"odinger, Dirac, and Pauli who were awarded Nobel Prizes for their theoretical work. Interestingly, the quantum-mechanical equation for the electron proposed by Pauli did not achieve the same widespread adoption as the Schr\"odinger and Dirac equations; Pauli instead received the Nobel Prize for his exclusion principle, according to which no two electrons can occupy the same quantum state. In Pauli's formulation, the electron's state is characterized by four quantum numbers, including a new quantum number that he introduced. This additional degree of freedom was identified in 1925 by George Uhlenbeck and Samuel Goudsmit as the electron spin.

A timely reminder of the \emph{centennial} of quantum theory is the recent review~\cite{Real-7}, which catalogues 39 distinct derivations of the Dirac equation and analyses them from a modern point of view. However, the author of~\cite{Real-7} candidly acknowledges that several derivations may have been overlooked, given the sheer number of approaches that exist. Indeed,  review~\cite{Real-7} does not devote sufficient attention to the formative period of 1925--1928, particularly the roles of such distinguished scientists as Wolfgang Pauli, Charles Galton Darwin, and Hendrik Kramers. The aim of the present article is to supplement review~\cite{Real-7}, primarily through a more detailed examination of the foundational period of quantum mechanics.

Furthermore, the year 2026 marks the centennial of the publication of a fundamental equation of relativistic quantum mechanics---the equation commonly known as the Klein--Gordon equation:
\begin{equation}\notag
\left(\frac{1}{c^{2}}\frac{\partial^{2}}{\partial t^{2}}-\nabla^{2}+\frac{m^{2}c^{2}}{\hbar^{2}}\right)\psi=0.
\end{equation}
This second-order relativistic wave equation, which describes spinless particles, occupies a central place in the development of quantum theory. It is, in fact, the relativistic generalization of the Schr\"odinger equation to which the latter reduces in the non-relativistic limit, and it served as the essential starting point from which Dirac derived his own first-order equation by seeking a ``square root'' of the Klein--Gordon operator. In this sense, the Dirac equation is directly and inseparably connected to the Klein--Gordon equation. The role of Erwin Schr\"odinger in this story is presented below near his equation ~(\ref{eq1}). 

On historical grounds, this equation would be more accurately called the \emph{Klein--Fock--Gordon equation}, a name that faithfully reflects the chronological order in which the three foundational papers appeared in \emph{Zeitschrift f\"ur Physik}: Oskar Klein~\cite{Klein1926} (Z.~Phys.\ \textbf{37}, 895, 1926), Vladimir Fock~\cite{Fock1926} (Z.~Phys.\ \textbf{39}, 226, received 30 July 1926), and Walter Gordon~\cite{Gordon1926} (Z.~Phys.\ \textbf{40}, 117, 1926). The designation Klein--Fock--Gordon is standard in the East European tradition, where Fock's priority has long been recognized. In the English-language tradition, however, Fock's name is regrettably omitted, and the equation is referred to simply as the Klein--Gordon equation. As emphasized in~\cite{Okun2010}, Fock was also the first to introduce the concept of an Abelian gauge transformation and to discover the gauge invariance of the electromagnetic interaction within the framework of quantum mechanics---contributions whose significance extends far beyond the equation itself. This scientist suggested a long list of contributions in the quantum field theory, which is only partially mention in the review \cite{Real-7}.

\section{Some Historical Context}

\subsection{Birth of a new theory}
Regarding quantum physics as a whole, it originates from the work~\cite{Real-8} of another Nobel laureate, Max Planck, and the theoretical models of Heisenberg, Schr\"odinger, Dirac, and Pauli rely not only on Max Planck's discovery but also on a whole series of works by no less famous authors such as Albert Einstein, Ernest Rutherford, Niels Bohr, Walther Gerlach and Otto Stern, Arthur Compton, and many others. The scientific breakthroughs of the remarkably productive year 1925 were made possible by the persistent and fruitful work of their predecessors from 1901 to 1924.

Even in a brief account of the history of the creation of quantum mechanics, it is necessary to mention the authors of the Klein--Gordon equation. It was from this equation that both Erwin Schr\"odinger and Paul Dirac derived their own equations.

Nevertheless, it should be noted that Erwin Schr\"odinger was aware of the second-order equation known today as the Klein--Gordon equation but did not publish it. He was searching for a version of this equation containing a first-order time derivative, with time serving as the evolution parameter. His guiding idea was the Hamiltonian approach. Accordingly, Schr\"odinger modified the equation, and his publication appeared somewhat later.

P.A.M.\ Dirac took into account both the Schr\"odinger and the Klein--Gordon equations. Reference~\cite{Real-7} illustrates the influence of the Dirac equation on one hundred years of development of relativistic quantum mechanics.

The year 2025 was proclaimed the International Year of Quantum Science and Technology in recognition of the centennial of Heisenberg's foundational publication~\cite{Real-1} and the theoretical breakthroughs that followed; see, for example, the editorial~\cite{Real-9}
in \emph{Physical Review Letters}.  That editorial identifies
three broad periods of quantum theory development, each rooted
in the work of a successive generation of theoretical physicists:
the foundational era discussed above, the period of consolidation
and expansion, and the modern era linked to quantum information
science.

In Subsection~2.3 and Section~3 below, we expand on the contribution of H.A.\ Kramers.

\subsection{Periods of Development}

The development period begins after the foundations have been laid and
agreement on the interpretation of quantum mechanics has been reached. It
encompasses the construction of quantum electrodynamics, quantum
chromodynamics, and the Standard Model, together with their successful
confrontation with experimental data. On this basis, a technological
breakthrough took place. Notable examples of quantum technologies include
lasers, magnetic resonance imaging, integrated circuits, and, more recently, quantum information processing and quantum sensing. Quantum theory also began to transform adjacent fields such as chemistry, materials science, astrophysics, and cosmology. This period was accompanied by a gradual
acceptance of the special effects that arise in the quantum regime.

Interestingly, review~\cite{Real-9} characterize the modern
period as one intimately connected with computer science. Having come to terms
with quantum strangeness, scientists realized that the quantum world possesses
an enormous, inherent capacity for information processing. The possibility of building a universal quantum computer and achieving a practical quantum computational advantage is particularly emphasized.

However, despite the predominantly positive developments, negative trends also
emerged. Confidence in the correctness of the Standard Model and of quantum
theory as a whole led to the appearance and rapid proliferation of a number of
purely theoretical models lacking experimental confirmation. A typical example
is the theory of supersymmetry. Nevertheless, the development of even such
speculative models simultaneously stimulated new mathematics and, in particular, new applied mathematics.

The successes of quantum physics and their technological consequences have gradually reshaped the prevailing philosophical outlook on the natural world---and, in particular, the philosophy of science and scientific methodology. The achievements of quantum physics have permeated everyday life, transforming not only technology but also the conceptual framework through which we understand physical reality.

\subsection{Overlooked Contributions}
 
Historians of physics have written dozens of books and hundreds of articles about the period of the ``quantum revolution.'' Unfortunately, however, historians are seldom theoretical physicists (with rare exceptions in which both professions are combined), and insufficient attention has been paid to the roles of scientists such as Charles Galton Darwin and Hendrik Kramers in laying the foundations of quantum mechanics.
 
In review~\cite{Real-7}, in particular, the role of Charles Galton Darwin (grandson of the even more famous Charles Robert Darwin) is highlighted. Darwin was the first to elucidate~\cite{Real-10} the details of the Dirac equation and its physical interpretation, and he proposed a quasi-relativistic approximation that includes a term now bearing his name. On this basis, Darwin developed his own approach to deriving the fine structure of the hydrogen atom from Dirac's relativistic electron equation. Moreover, the connection between the Dirac equation and Maxwell's equations that he identified is proposed in~\cite{Real-7} to be termed the \emph{Darwin vector}.
 
The birth of quantum mechanics was accompanied by a number of lively discussions, remarkable stories, and even tragedies---though the tragedies are not the subject of our discussion. Monograph~\cite{Real-11} describes a particularly illuminating situation that arose from discussions between Heisenberg and Dirac, as well as between Pauli and Kramers, concerning the incorporation of electron spin into the emerging quantum mechanics.
 
In what follows we draw upon Dresden's monograph~\cite{Real-11} through direct quotations together with some necessary commentary. ``It is illuminating that on February 5, 1927, Heisenberg wrote a letter to Pauli in which he mentions that he made a bet with Dirac that it would take 3 years to understand the spin fully. Dirac claimed this would be clear in 3 months.''
 
Shortly thereafter, Pauli published paper~\cite{Real-12} in which substantial progress was made in combining quantum mechanics with spin. Indeed, it was in this paper that the explicit matrix representation of the electron spin in the form of $2\times 2$ matrices was proposed for the first time. ``Of course he does not explain the spin, he just incorporates its description in quantum terms. This was an important advance, albeit a partial one. Heisenberg mentions the bet with Dirac in a letter to Jordan (May 1927); he feels that the insufficiencies in Pauli's paper will guarantee that he will win his bet. Pauli and others were dissatisfied with Pauli's paper because it did not meet the requirements of relativity. In fact, Pauli claimed that a consistent relativistic spin theory was in principle impossible.''
 
Further, according to~\cite{Real-11}: ``Kramers did not believe this; so Kramers and Pauli made a bet that one could not construct a relativistic spin theory. Pauli issued a direct challenge to this effect to Kramers. Both Dirac and Kramers lost their bets---but Dirac was certainly the moral victor and Kramers demonstrated eventually, even to Pauli's satisfaction, that he had effectively met Pauli's challenge'' (\cite{Real-11}, p.~64).
 
Dirac published his famous paper~\cite{Real-5} approximately six months after the bet with Heisenberg expired, producing a formalism that satisfied both the requirements of relativity and quantum mechanics. ``At about that time (January 1928), Kramers had obtained, by a very intricate, complicated, and somewhat contrived method, a set of equations that had exactly the same mathematical and physical content as Dirac's equation. (It actually takes some work to demonstrate that precise equivalence.) Kramers was very disappointed that even though he (Kramers) had achieved his goal, Dirac was ahead of him by just a few months.'' Dresden continues: ``Perhaps because his derivation was somewhat cumbersome and Dirac's derivation was so elegant, he did not publish any of this until about 10 years later.'' An additional reason may have been his discussions with W.~Pauli.
 
Pauli's skepticism was strikingly evident in his reaction to Kronig's report on the discovery of electron spin. He categorically opposed Kronig's ideas and proposals, thereby delaying R.L.~Kronig's publication; in the interim, an article by other authors appeared. Today, we associate the discovery of spin with G.E.~Uhlenbeck and S.~Goudsmit. In fairness, it should be noted that Kronig was likewise unsupported at the time by Heisenberg and Kramers~\cite{Real-13}.
 
Hendrik Anthony Kramers was, during this period, the assistant of Niels Bohr. A man of great modesty in character yet a towering theoretical physicist, Kramers derived the fundamental equation of relativistic quantum mechanics independently of, and essentially simultaneously with, Paul Dirac---but did not publish his result  in a timely fashion. At that time, many scientists tested and verified their results through correspondence with Wolfgang Ernst Pauli, who was an avid correspondent but who, over the years, became renowned for his sharp skepticism. Pauli was skeptical of Kramers' results as well, and under the influence of Pauli's authority, Kramers delayed his publication by seven years~\cite{Real-14}.
 
The story of Kramers' parallel derivation was well known to contemporaries and is described in monographs~\cite{Real-11,Real-15}. Only several years after the success and recognition of Dirac's result~\cite{Real-5} did Pauli reconsider Kramers' proof and sanction it for publication~\cite{Real-14,Real-16}. Pauli remained highly critical, but after lengthy correspondence he conceded that Kramers' procedure was legitimate, though ``not elegant.''
 
As viewed from today's perspective, the derivation possesses certain attractive features, chief among them the natural relativistic invariance of the proof. It is also noteworthy that Kramers' approach in no way appeals to ``extracting the square root of the Klein--Gordon operator.'' Furthermore, Kramers' approach influenced several other derivations reviewed in~\cite{Real-7}.

The authors of monographs~\cite{Real-11,Real-15} were intimately familiar with Kramers' scientific legacy and personal history, and their accounts preserve the story of the wager among several founders of quantum theory. Kramers numbered among those who lost the bet; whether any of the participants ever settled the stakes remains, to this day, unknown.
  
Thus, the list of 39 derivations of the Dirac equation compiled in~\cite{Real-7} can be significantly supplemented, not only by Kramers' results~\cite{Real-14,Real-16} but also by the Van der Waerden approach  and recent derivations.

\section{A few forgotten methods from the birth of quantum mechanics}

A rough classification of the methods for deriving the Dirac equation surveyed in~\cite{Real-7} reveals two broad groups of approaches: those that appeal to the factorization of the Klein--Gordon equation, and those that proceed by other means. In what follows, we concentrate on the second group.

\subsection{The Van der Waerden Derivation}

Review~\cite{Real-7} noted the Van der Waerden--Sakurai method of deriving the Dirac equation, which Sakurai subsequently popularized in his well-known monographs. Since, however, Van der Waerden's own derivation was not presented in~\cite{Real-7}, we comment here on his contributions to the Dirac theory, published in~\cite{Real-17} and in his monograph~\cite{Real-18}---both of which appeared before Kramers' result~\cite{Real-14}, which became known only in 1935.

We begin with Van der Waerden's derivation because of an important---if passing---reference to Van der Waerden in Kramers' article~\cite{Real-14}. This reference raises natural questions: to what extent did Van der Waerden's work influence~\cite{Real-14}, and was Kramers' derivation, presumably completed in 1928 at essentially the same time as Dirac's, truly independent of Van der Waerden's results?

Partial answers can be found in~\cite{Real-14,Real-17,Real-18}. We do not, however, presume to adjudicate this matter. Our goal is rather to highlight Kramers' role and to draw attention to his contribution to the birth of quantum theory and, more broadly, to the entire formative period of quantum mechanics. Detailed questions concerning the interval 1928--1934 shall be left to historians of physics.

Like Dirac's original derivation~\cite{Real-5}, the approach taken in~\cite{Real-18} is based on factorization of the Klein--Gordon operator. There is, however, a significant difference that renders Van der Waerden's treatment both independent and genuinely insightful.

The principal steps parallel those in~\cite{Real-5} but are carried out at the two-component level. Moreover, Van der Waerden appealed fundamentally to the principle of relativistic invariance, thereby introducing the group-theoretical approach to the problem. Within this framework the Klein--Gordon equation was referred to as the relativistic Schr\"odinger equation. Recall that Erwin {Schr\"odinger} was the first {to derive} the equation {that} we call today {the} second{-}order Klein--Gordon equation. Nevertheless, in the paper~\cite{Real-4} he {abandoned} the second{-}order equation and modified it {into a} non{-}relativistic but very useful form. The guiding principle was the requirement {that a} transition to the stationary problem {be possible:}
\begin{equation}
\label{eq1}
\hat{H}\psi = E\psi.
\end{equation}
Note that for the Klein--Gordon equation such a transition is impossible.

This principle was {also} the guiding one for {Van der Waerden, who} started from the relativistic {Schr\"odinger} equation
\begin{equation}
\label{eq2}
(c^{-2}d^{2}_{t}-d^{2}_{x}-d^{2}_{y}-d^{2}_{z})\psi = \mu^{2}c^{2}\psi,
\end{equation}
where
\begin{equation}
\label{eq3}
d_{t}=-i\hbar \frac{\partial}{\partial t}-e\varphi, \quad d_{x,y,z}=p_{x,y,z}+\frac{e}{c}A_{x,y,z},
\end{equation}
and
\begin{equation}
\label{eq4}
p_{x}=\frac{\hbar}{i}\frac{\partial}{\partial x}, \quad p_{y}=\frac{\hbar}{i}\frac{\partial}{\partial y}, \quad p_{z}=\frac{\hbar}{i}\frac{\partial}{\partial z}.
\end{equation}
Here $\varphi$ is the scalar (electric) potential and $\vec{A}$ is {the} vector {potential} (called in~\cite{Real-18} {the} magnetic potential).

His next independent step was {to take} into account {the} electron spin and {to consider a} wave equation for {a} two{-}component wave function, where the spin was naturally related to the standard $2\times 2$ Pauli matrices.

Further, the operator in {the corresponding} relativistic {Schr\"odinger} equation is factorized into two factors{,} and the {resulting} equation is given by
\begin{equation}
\label{eq5}
(c^{-1}d_{t}-d_{x}\sigma_{x}-d_{y}\sigma_{y}-d_{z}\sigma_{z})(c^{-1}d_{t}+d_{x}\sigma_{x}+d_{y}\sigma_{y}+d_{z}\sigma_{z})\Psi=\mu^{2}c^{2}\Psi.
\end{equation}
This factorization is valid only if all operators $d_{t},d_{x},d_{y},d_{z}$ commute {with one another,} and coincides with equations \eqref{eq2} and \eqref{eq3} only for constant potentials $\varphi$ and $\vec{A}$.

After that, Van der Waerden considered the general case of non{-}constant potentials and proved{,} after some algebraic transformations{,} that in {this} formalism equation (5) is equivalent {to a} pair of equations for two different 2-component wave functions{:}
\begin{equation}
\label{eq6}
(\frac{1}{c}d_{t}+d_{x}\sigma_{x}+d_{y}\sigma_{y}+d_{z}\sigma_{z})\Psi=-\mu c \dot{\Psi}, \quad (\frac{1}{c}d_{t}-d_{x}\sigma_{x}-d_{y}\sigma_{y}-d_{z}\sigma_{z})\dot{\Psi}=-\mu c\Psi.
\end{equation}
This is the natural transition from the two-component to the four-component formalism.

In addition{,} within the framework of the group-theoretical approach proposed by him in the previous chapters of~\cite{Real-18}, van der Waerden notes that the introduction of an additional function $\dot{\Psi}$ is necessary since the wave equation should be invariant not only under the proper Lorentz transformation, but also under {space--time} reflections (the complete group). Moreover, for the 4-component wave function the equation
\begin{equation}
\label{eq7}
i\hbar\frac{\partial}{\partial t}\Psi=\hat{H}\Psi
\end{equation}
is valid, where $\hat{H}$ is {a} linear self-adjoint operator. Therefore, in the stationary case~\eqref{eq1} we have real eigenvalues.

Instead of four components $\Psi, \dot{\Psi}$ it is {convenient} to introduce {four other} components $\Psi^{r}_{\lambda}, \Psi^{a}_{\lambda}${:}
\begin{equation}
\label{eq8}
\Psi^{r}_{\lambda}=\Psi_{\lambda}+\dot{\Psi}^{\lambda}, \quad \Psi^{a}_{\lambda}=\Psi_{\lambda}-\dot{\Psi}^{\lambda},\quad (\lambda=1,2).
\end{equation}
Within the framework of his group-theoretical approach, Van der Waerden commented on this transition. Under rotations, ($\Psi^{r}_{\lambda}, \dot{\Psi}_{\lambda}^{a}$) transform in the same way as ($\Psi_{\lambda}, \dot{\Psi}^{\lambda}$), whereas under reflections $r$ the transformation law takes the form
\begin{equation}
\label{eq9}
r\Psi^{r}_{\lambda}=\Psi_{\lambda}^{r}, \quad r\Psi^{a}_{\lambda}=-\Psi_{\lambda}^{a}.
\end{equation}

Hence, after {summation, subtraction,} and multiplication by $c$ we {pass} from the system of equations \eqref{eq6} to the equations
\begin{equation}
\label{eq10}
(dt+\mu c^{2})\Psi^{r}+c(d_{x}\sigma_{x}+d_{y}\sigma_{y}+d_{z}\sigma_{z})\Psi^{a}=0, \, (dt-\mu c^{2})\Psi^{a}+c(d_{x}\sigma_{x}+d_{y}\sigma_{y}+d_{z}\sigma_{z})\Psi^{r}=0,
\end{equation}
which constitute the Dirac equation in its two-component form.

In~\cite{Real-5} this equation was introduced in a more attractive and compact 4-component spinor form. Nevertheless{,} the first group-theoretical derivation of Van der Waerden was very useful. Its subsequent influence on Sakurai, Ryder, and Kramers is evident.  We reiterate that Kramers in~\cite{Real-14} referred to Van der Waerden.

\subsection{Kramers derivation}

\subsubsection{Classical spin problems}

The central idea of Kramers's approach~\cite{Real-14} consists in formulating a Hamiltonian description of the classical spin of the electron and then quantizing it. As a consequence, the Thomas precession factor is reached. 

Kramers began in~\cite{Real-14} from the equation governing the precession of the spin-vector, written in relativistically invariant form:
\begin{equation}
\label{eq11}
\frac{d\textbf{S}}{d\tau}=\alpha [\textbf{S}\textbf{F}].
\end{equation}
Here $d\tau$ denotes the element of proper time, whereas \textbf{S} and \textbf{F} are two complex vectors given by
\begin{equation}
\label{eq12}
\textbf{S}=\textbf{A}+i\textbf{B}, \quad \textbf{F}=\textbf{H}+i\textbf{E}.
\end{equation}
The real vectors \textbf{A} and \textbf{B}, which characterize the spin, transform under a Lorentz transformation in the same way as \textbf{H} and \textbf{E} (the magnetic and electric field strengths); $[\textbf{SF}]$ denotes the vector product, and the constant~$\alpha$, as shown below, turns out to be the ratio of the magnetic moment to the angular momentum.

The assumption that \textbf{B} always vanishes in the instantaneous rest frame of the electron leads to the relativistically invariant relation
\begin{equation}
\label{eq13}
\textbf{B}=\frac{1}{c}[\textbf{A}\textbf{v}],
\end{equation}
where \textbf{v} is the velocity of the electron. A further assumption is that, in the frame where $\textbf{v}=0$, the real part of~(\ref{eq11}) reduces to the non-relativistic classical equation for a spinning electron with spin-vector~\textbf{A} (i.e.\ the angular-momentum vector) in a magnetic field~\textbf{H}:
\begin{equation}
\label{eq14}
\dot{\textbf{A}}=\alpha[\textbf{A}\textbf{H}].
\end{equation}
The imaginary part of~(\ref{eq11}) reduces in the same frame to the expression
\begin{equation}
\label{eq15}
\dot{\textbf{B}}=\alpha[\textbf{A}\textbf{E}].
\end{equation}
When the reaction of the spin on the orbital motion is very small, the orbital equation of motion (still for $\textbf{v}=0$) takes the form
\begin{equation}
\label{eq16}
\dot{\textbf{v}}m=-e\textbf{E}.
\end{equation}
Substituting into~(\ref{eq15}) yields
\begin{equation}
\label{eq17}
\dot{\textbf{B}}=-\frac{\alpha m}{e}[\textbf{A}\dot{\textbf{v}}].
\end{equation}
A comparison of~(\ref{eq17}) with the time derivative of~(\ref{eq13}) at $\textbf{v}=0$ immediately gives
\begin{equation}
\label{eq18}
\alpha = -\frac{e}{mc},
\end{equation}
in agreement with the Thomas precession factor of $1/2$.

Having established this result, Kramers noted that a more complete classical theory remained to be developed but did not pursue the matter further. He employed this formalism only as a stepping stone toward quantum theory and, ultimately, toward a derivation of the Dirac equation.

\subsubsection{Canonical Hamilton formalism}

Before quantizing equations~(\ref{eq11}) and~(\ref{eq16}), Kramers sought a single Hamiltonian from which both could be simultaneously derived. To this end, he first considered equation~(\ref{eq14}), which involves only real vectors. This equation can be cast in canonical form if, in agreement with the standard treatment of a dipole in an external field, the energy is taken to be
\begin{equation}
\label{eq19}
H_{A}=-\alpha (\textbf{A}\textbf{H}).
\end{equation}

There is only one degree of freedom, and for the canonical coordinates one may choose $p=A_{1}, \, q=\arctan\frac{A_{2}}{A_{3}}, \, (A_{2}+iA_{3}=\sqrt{A^{2}-p^{2}}e^{iq})$. The corresponding Poisson brackets take the form $\{A_{1}A_{2}\}=\frac{\partial A_{1}}{\partial p}\frac{\partial A_{2}}{\partial q}-\frac{\partial A_{1}}{\partial q}\frac{\partial A_{2}}{\partial p}=-A_{3}, \, \{A_{2}A_{3}\}=-A_{1}, \, \{A_{3}A_{1}\}=-A_{2}$. The equations $A_{k}=-\sum_{\ell}\{A_{k}A_{\ell}\}\frac{\partial H_{A}}{\partial A_{\ell}}$, governing the evolution of~\textbf{A}, are then seen to be identical to~(\ref{eq14}) once the expression~(\ref{eq19}) for~$H_{A}$ is adopted.

The equations~(\ref{eq11}) can be treated in a similar fashion
\begin{equation}
\label{eq20}
H_{S}=-\alpha (\textbf{S}\textbf{F}), \quad \{S_{k}S_{\ell}\}=-\varepsilon^{k\ell m}S_{m}, \, \varepsilon^{123}=+1, \quad \frac{d S_{k}}{d \tau}=-\sum_{\ell}\{S_{k}S_{\ell}\}\frac{\partial H_{S}}{\partial S_{\ell}}.
\end{equation}
where $\varepsilon^{k\ell m}$ is the completely antisymmetric Levi-Civita tensor.

Formally, one still has a system of one degree of freedom, the canonical coordinates in this case being, e.g., $p=S_{1}, \, q=\arctan \frac{S_{2}}{S_{3}}$. Here, however, these coordinates are complex, and so is the Hamiltonian~$H_{S}$. Separating real and imaginary parts---$p=p^{\prime}+ip^{\prime\prime}, \, q=q^{\prime}-iq^{\prime\prime}, \, H=H^{\prime}(p^{\prime}p^{\prime\prime}q^{\prime}q^{\prime\prime})+iH^{\prime\prime}(p^{\prime}p^{\prime\prime}q^{\prime}q^{\prime\prime})$---it is readily verified that the complex equations of motion $\frac{d p}{d\tau}=-\frac{\partial H}{\partial q}, \, \frac{d q}{d\tau}=\frac{\partial H}{\partial p}$ with real~$d\tau$ correspond to real canonical equations of a system with two degrees of freedom, in which either $H^{\prime}$ or $H^{\prime\prime}$ serves as the Hamiltonian:
\begin{equation}
\label{eq21}
\frac{d p^{\prime}}{d\tau}=-\frac{\partial H^{\prime}}{\partial q^{\prime}}, \, \frac{d q^{\prime}}{d\tau}=\frac{\partial H^{\prime}}{\partial p^{\prime}}, \, \frac{d p^{\prime}}{d\tau}=-\frac{\partial H^{\prime\prime}}{\partial q^{\prime\prime}}, \, \frac{d q^{\prime\prime}}{d\tau}=-\frac{\partial H^{\prime\prime}}{\partial p^{\prime}},
\end{equation}
or
$$\frac{d p^{\prime\prime}}{d\tau}=-\frac{\partial H^{\prime}}{\partial q^{\prime\prime}}, \, \frac{d q^{\prime\prime}}{d\tau}=\frac{\partial H^{\prime}}{\partial p^{\prime\prime}}, \, \frac{d p^{\prime\prime}}{d\tau}=-\frac{\partial H^{\prime\prime}}{\partial q^{\prime}}, \, \frac{d q^{\prime}}{d\tau}=-\frac{\partial H^{\prime\prime}}{\partial p^{\prime\prime}}.$$

This analysis reveals that, besides~(\ref{eq20}), two alternative routes to deriving the equations of motion from a real Hamiltonian are available. To this end, one introduces, alongside~\textbf{S}, its complex conjugate~$\textbf{S}^{*}$:
$$H_{S}=-\alpha \frac{(\textbf{S}\textbf{F})+(\textbf{S}^{*}\textbf{F}^{*})}{2},$$
$$ \quad \quad \{S_{k}S_{\ell}\}=-2\varepsilon^{k\ell m}S_{m}, \quad \{S^{*}_{k}S^{*}_{\ell}\}=-2\varepsilon^{k\ell m}S^{*}_{m}, \quad \{S_{k}S^{*}_{\ell}\}=0, \quad \quad \quad \quad   (20a)$$

$$H_{S}=-\alpha \frac{(\textbf{S}\textbf{F})-(\textbf{S}^{*}\textbf{F}^{*})}{2i},$$
$$ \quad \quad \{S_{k}S_{\ell}\}=-2i\varepsilon^{k\ell m}S_{m}, \quad \{S^{*}_{k}S^{*}_{\ell}\}=-2i\varepsilon^{k\ell m}S^{*}_{m}, \quad \{S_{k}S^{*}_{\ell}\}=0, \quad \quad \quad \quad   (20b)$$
$$\frac{d S^{(*)}_{k}}{d \tau}=-\sum_{\ell}\{S^{(*)}_{k}S_{\ell}\}\frac{\partial H_{S}}{\partial S_{\ell}}-\sum_{\ell}\{S^{(*)}_{k}S^{*}_{\ell}\}\frac{\partial H_{S}}{\partial S^{*}_{\ell}},$$
where $S^{(*)}_{k}$ denotes either $S_{k}$ or $S^{*}_{k}$.

The system is now explicitly treated as one with two degrees of freedom. The Poisson brackets are obtained by taking the four real canonical variables [cf.~(\ref{eq21})] explicitly into account.

For completeness, one may also mention the alternative to~(\ref{eq20}) that arises when the complex conjugate~$\textbf{S}^{*}$ is used in place of~\textbf{S}:
$$H^{*}_{S}=-\alpha (\textbf{S}^{*}\textbf{F}^{*}), \quad \{S^{*}_{k}S^{*}_{\ell}\}=-\varepsilon^{k\ell m}S^{*}_{m}, \quad \frac{d S^{*}_{k}}{d \tau}=-\sum_{\ell}\{S^{*}_{k}S^{*}_{\ell}\}\frac{\partial H^{*}_{S}}{\partial S^{*}_{\ell}}. \quad \quad \quad (20c)$$

The relation $(\textbf{S}\textbf{F})=\{(\textbf{A}\textbf{H})-(\textbf{B}\textbf{E})\}+i\{(\textbf{A}\textbf{E})+(\textbf{B}\textbf{H})\}$ follows from~(\ref{eq12}).

It would therefore appear most natural to adopt the form~$H^{\prime}_{S}$ given by~(20a), namely $H^{\prime}_{S}=-\alpha \{(\textbf{A}\textbf{H})-(\textbf{B}\textbf{E})\}$, since in the instantaneous rest frame of the electron it reduces to the familiar real energy expression~(\ref{eq19}). If one adopts instead the simpler form~$H_{S}$ or~$H^{*}_{S}$ given by~(20) or~(20c), it appears at first sight as though one has formally introduced an imaginary electric moment of the electron equal to $-i\alpha \textbf{A}$ or $+i\alpha \textbf{A}$, respectively.

The equations of motion~(\ref{eq16}), in which the spin is neglected, can be derived from the familiar Hamiltonian
\begin{equation}
\label{eq22}
H_{0}\equiv -\frac{(\epsilon + e\Phi)^{2}}{c^{2}}+(\textbf{p}+\frac{e}{c}\Psi)^{2}=-\frac{1}{2}mc^{2},
\end{equation}
$$\frac{d\epsilon}{d\tau}=\frac{\partial H_{0}}{\partial t}, \, \frac{d t}{d\tau}=-\frac{\partial H_{0}}{\partial \epsilon}, \, \frac{d p_{i}}{d\tau}=-\frac{\partial H_{0}}{\partial x_{i}}, \, \frac{d x_{i}}{d\tau}=\frac{\partial H_{0}}{\partial p_{i}},$$
where $\epsilon$ is the energy and \textbf{p} the momentum of the system, $\Phi$~is the scalar potential, and $\Psi$~is the vector potential of the external field. Note that in modern literature the three-vector potential of the external electromagnetic field is denoted by~\textbf{A} and, for a general external field, by~\textbf{V}. Here the symbol~\textbf{A} carries a different meaning.

Within the limits of validity of the classical analysis (i.e.\ \textbf{S} so small that the reaction of the spin on the orbital motion is negligible), a Hamiltonian~$\overline{H}$ that simultaneously governs the orbital motion and the spin precession is obtained simply by adding the spin Hamiltonian to~$H_{0}$ in~(\ref{eq22}):
$$\overline{H}\equiv \frac{1}{2m}\left\{-\frac{(\epsilon + e\Phi)^{2}}{c^{2}}+(\textbf{p}+\frac{e}{c}\Psi)^{2}\right\}+\frac{e}{mc}(\textbf{S}\textbf{F})=-\frac{1}{2}mc^{2}.$$
Multiplying through by~$-2m$, Kramers arrived at the Hamiltonian
\begin{equation}
\label{eq23}
H \equiv \left\{\frac{(\epsilon + e\Phi)^{2}}{c^{2}}-(\textbf{p}+\frac{e}{c}\Psi)^{2}\right\}-\frac{e}{mc}(\textbf{S}\textbf{F})=m^{2}c^{2},
\end{equation}
Here the formulation is based on~(\ref{eq20}). Using instead~(20a),~(20b), or~(20c), one may replace $(\textbf{S}\textbf{F})$ in~(\ref{eq23}) by $\frac{1}{2}\{(\textbf{S}\textbf{F})+(\textbf{S}^{*}\textbf{F}^{*})\}$, $-\frac{1}{2}i\{(\textbf{S}\textbf{F})-(\textbf{S}^{*}\textbf{F}^{*})\}$, or $(\textbf{S}^{*}\textbf{F}^{*})$, respectively. Given the approximation involved (\textbf{S} very small), any of these four expressions may be chosen, although $\frac{1}{2}\{(\textbf{S}\textbf{F})+(\textbf{S}^{*}\textbf{F}^{*})\}$ might seem the most natural.

Kramers observed that it appears difficult, if not impossible, to construct a Hamiltonian formulation in which the constraint~(\ref{eq13})---which has the practical effect of reducing the two spin degrees of freedom to one---is automatically satisfied.

He further explained that this question cannot be settled before a classical system of equations of motion has been established in which the condition~(\ref{eq13}) is rigorously fulfilled, rather than only approximately as in the Hamiltonians considered above.

\subsubsection{Quantization}

On this basis, Kramers carried out the quantum-mechanical quantization of the system. To quantize the motion governed by the Hamiltonian~(\ref{eq23}) or one of its alternatives, $H$ must be promoted to an \emph{operator}~$H_{op}$ acting on a wave function~$\psi$:
\begin{equation}
\label{eq24}
H_{op}\psi = m^{2}c^{2}\psi.
\end{equation}

If the spin quantum number is to take the value~$\frac{1}{2}$, and if one follows Kramers in restricting attention to~(\ref{eq23}), then the following familiar operator representations must be adopted:
\begin{equation}
\label{eq25}
\epsilon =ih\frac{\partial}{\partial t}, \quad p_{i}=-ih\frac{\partial}{\partial x_{i}},
\end{equation}
\begin{equation}
\label{eq26}
S_{1} = \frac{h}{2}\left| {{\begin{array}{*{20}c}
 1 \hfill &  0 \hfill\\
 0 \hfill & -1  \hfill\\
\end{array} }} \right|, \quad S_{2}=\frac{h}{2}\left| {{\begin{array}{*{20}c}
 0 \hfill &  1 \hfill\\
 1 \hfill &   0  \hfill\\
 \end{array} }} \right|, \quad S_{3}=\frac{h}{2}\left| {{\begin{array}{*{20}c}
 0 \hfill &  -i \hfill\\
 i \hfill &   0  \hfill\\
 \end{array} }} \right|.
\end{equation}
The commutation properties of these expressions satisfy the conditions that correspond to the analogous Poisson brackets. The introduction of the Pauli spin matrices for~\textbf{S} means that $\psi$, in addition to its dependence on~$x_{1},x_{2},x_{3},t$, depends on a spin variable that can take only two values---for instance, the eigenvalues $\pm \frac{h}{2}$ of~$S_{1}$---so that $\psi$ can be represented as a pair of wave components $\psi_{+}$ and~$\psi_{-}$.

Note that the matrices in~(\ref{eq26}) coincide with those introduced by Pauli; only the notation $S_{1}, \, S_{2}, \, S_{3}$ is somewhat unusual by modern standards. The corresponding Dirac gamma matrices can, of course, be chosen in any unitarily equivalent form, the sole requirement being that the anticommutation relations of the Clifford algebra generators be satisfied. The essentially distinct possibilities for such a choice are discussed in~\cite{Real-19}.

The relativistic invariance of this representation for~\textbf{S} was established by Weyl and van der Waerden~\cite{Real-17,Real-18}, who showed that to each Lorentz transformation one can assign a unimodular transformation of the wave components such that $\psi^{2}_{+}, \, \psi_{+}\psi_{-}$, and $\psi_{-}$ transform as $-F_{2}+iF_{3}, \, F_{1}$, and $F_{2}+ iF_{3}$, respectively. Under this convention the Pauli matrix components transform in exactly the same way as the components of $\textbf{F} = \textbf{H}+i\textbf{E}$. Kramers further noted that the components of~\textbf{S} remain Hermitian even when subjected to a complex orthogonal transformation.

Substituting equations~(\ref{eq25}) and~(\ref{eq26}) into~(\ref{eq23}) to construct~$H_{op}$, equation~(\ref{eq24}) takes precisely the form that Dirac obtained by ``squaring'' his linear equations. The latter can be recovered from~(\ref{eq24}) by observing that $H_{op}$, as thus constructed, factorizes as follows:
$$H_{op} \equiv \left\{\frac{(\epsilon + e\Phi)}{c}-(\textbf{p}+\frac{e}{c}\Psi , \textbf{S})\right\}\left\{\frac{(\epsilon + e\Phi)}{c}+(\textbf{p}+\frac{e}{c}\Psi , \textbf{S})\right\}$$
Consequently, defining
\begin{equation}
\label{eq27}
\left\{\frac{(\epsilon + e\Phi)}{c}+(\textbf{p}+\frac{e}{c}\Psi , \textbf{S})\right\}\psi =mc\chi,
\end{equation}
one finds that $\chi$ satisfies
\begin{equation}
\label{eq28}
\left\{\frac{(\epsilon + e\Phi)}{c}-(\textbf{p}+\frac{e}{c}\Psi , \textbf{S})\right\}\chi =mc\psi.
\end{equation}
Since $\psi$ and $\chi$ are both two-component wave functions, equations~(\ref{eq27}) and~(\ref{eq28}) constitute a system of four equations, which are equivalent to the Dirac equations.

\subsubsection{Concluding comments on the Kramers derivation}

The derivation presented above constitutes a genuinely independent route to the Dirac equation. Nevertheless, certain questions of priority and influence deserve consideration. The principal one concerns the extent to which Kramers drew upon van der Waerden's results, which are explicitly cited in his paper~\cite{Real-14} and in the discussion above. Indeed, at the close of~\cite{Real-14}, Kramers again invoked van der Waerden~\cite{Real-17,Real-18}, noting that the transformation properties of the two-component wave functions---specifically, that $\chi^{2}_{+}, \chi_{+}\chi_{-}$ and $\chi^{2}_{-}$ transform like $-\psi^{*2}_{-}, \psi^{*}_{-}\psi^{*}_{+},-\psi^{*2}_{+}$, so that $\chi_{+}$ and $\chi_{-}$ transform like $\psi^{*}_{-}$ and $-\psi^{*}_{+}$ respectively---were already well established through van der Waerden's analysis.

If one accepts that Kramers completed his derivation in 1928, only a few weeks after Dirac, the question naturally arises as to what form it would have assumed in the absence of van der Waerden's group-theoretical framework. This remains, of course, a matter of speculation; what can be said with certainty is that the two lines of investigation---Kramers' physical, Hamiltonian-based approach and van der Waerden's algebraic one---were deeply complementary.

\subsection{Derivation via Operational Dynamical Modeling}

In~\cite{Real-20}, a conceptually distinctive route to the Dirac equation is provided by the framework of Operational Dynamical Modeling (ODM)~\cite{Real-21} and extended to the relativistic domain~\cite{Real-22}. ODM deduces equations of motion from two ingredients: (i) empirically motivated Ehrenfest-like relations governing the evolution of observable averages, and (ii) a specification of the algebra of observables. The Dirac equation then emerges as the unique dynamical model compatible with relativistic Ehrenfest relations and the canonical commutation relation between position and momentum. Moreover, imposing commutativity of position and momentum instead—together with the requirement that no antiparticles be created—yields the classical Spohn equation~\cite{Real-23}, which is thereby identified as the Koopman–von Neumann counterpart of the Dirac equation.

\subsubsection{Spinorial formulation of classical mechanics}

The starting point is the observation that relativistic classical mechanics admits a natural spinorial formulation~\cite{Real-24,Real-25,Real-26}. The proper velocity of a classical particle is encoded in the Feynman slash notation as
\begin{equation}
\label{eq29}
   u\!\!\!/ =  u^{\mu}\gamma_{\mu} = u_{\mu} \gamma^{\mu},
\end{equation}
where the gamma matrices satisfy the Clifford algebra
$(\gamma^{\mu}\gamma^{\nu} +\gamma^{\mu}\gamma^{\nu}) = 2g^{\mu \nu} \mathbf{1}$
with the metric signature $g_{\mu\nu} = \mathrm{diag}(1,-1,-1,-1)$.
A restricted Lorentz transformation is represented by a spinor
$L \in \mathbf{Spin}_{+}(1,3)$, and the leftmost column of $L$ is precisely a Dirac
four-component spinor~$\Psi$.

In this language, the classical equations of motion take a form that closely
resembles relativistic Ehrenfest relations:
\begin{equation}
\label{eq30}
 \frac{d x^{\mu}}{d\tau} = \Psi^\dagger c \gamma^0\gamma^{\mu}\Psi, \qquad
 \frac{d p_{\mu} }{d \tau}   =  c e \Psi^{\dagger}   \gamma^0  (\partial_{\mu}  A \!\!\!/ \,) \Psi,
\end{equation}
where $\tau$ is the proper time and the normalization $\Psi^\dagger\Psi = 1$ is imposed.
Crucially, these expressions are purely classical; the gamma matrices serve only to extract
the velocity stored in the spinor. The resemblance to quantum expectation values is what
makes the ODM program possible.

\subsubsection{Deduction of the Dirac Hamiltonian}

ODM requires three inputs to construct a dynamical mode~\cite{Real-20}. The evolution of average values in the form of Ehrenfest-like relations. (ii) The definition of the observables' average (the mathematical representation of $\langle \cdots \rangle$). (iii) The algebra of the observables.

Converting from proper time to coordinate time via $d/d\tau = \gamma\, d/dt$, the classical
spinorial equations motivate the following \emph{postulated} Ehrenfest-like relations for
relativistic dynamics:
\begin{equation}
\label{eq31}
 \frac{d}{d t}\langle\hat{\mathbf{x}}^{k}\rangle = \langle c \gamma^{0}\gamma^{k} \rangle, \quad \frac{d}{d t}\langle\hat{\mathbf{p}}_{k}\rangle=\langle ce\partial_{k}\hat{A}_{\nu}\gamma^{0}\gamma^{\nu} \rangle,
\end{equation}
where $\langle \cdots \rangle$ denotes an empirical average.
For input (ii), the averages are represented by the Dirac bra-ket inner product,
$\langle \cdots \rangle = \langle \psi | \cdots | \psi \rangle$, in a spinorial Hilbert space.

By Stone's theorem, unitary evolution in the Hilbert space implies the existence of a
self-adjoint generator $H$ such that
\begin{equation}
\label{eq32}
 i \hbar \frac{d |\psi \rangle}{dt}= H |\psi \rangle .
\end{equation}
Substituting this into the Ehrenfest relations and requiring validity for all initial states yields
the operator identities
\begin{equation}
\label{eq33}
\frac{1}{i\hbar}[\hat{\mathbf{x}}^{k},H]=c\gamma^{0}\gamma^{k}, \quad \frac{1}{i\hbar}[\hat{\mathbf{p}}_{k},H] = ce\partial_{k} \hat{A}_{\nu}\gamma^{0}\gamma^{\nu}.
\end{equation}
For input (iii), one specifies the canonical commutation relations
\begin{equation}
\label{eq34}
[\hat{\mathbf{x}}^{j},\hat{\mathbf{p}}_{k}]= -i\hbar \delta^{j}_{\,\, k}.
\end{equation}
Assuming $H = H(\hat{\mathbf{x}}^{k},\hat{\mathbf{p}}_{k})$, the commutator identities
reduce to a system of partial differential equations for the unknown~$H$:
\begin{equation}
\label{eq35}
-\frac{\partial}{\partial \hat{\mathbf{p}}_{k}}H=c\gamma^{0}\gamma^{k}, \quad \frac{\partial}{\partial \hat{\mathbf{x}}^{k}}H = ce\partial_{k} \hat{A}_{\nu}\gamma^{0}\gamma^{\nu}.
\end{equation}
Direct integration yields
\begin{equation}
\label{eq36}
 H(\hat{\mathbf{x}}^{k},\hat{\mathbf{p}}_{k})=-c\gamma^{0}\gamma^{k}\hat{\mathbf{p}}_{k}+ ce\hat{A}_{\nu}\gamma^{0}\gamma^{\nu}+C,
\end{equation}
where $C$ is a constant matrix. Requiring that $H$ reproduces the correct classical
Hamiltonian in the commutative limit fixes $C = mc^2 \gamma^0$~\cite{Real-22}.
The resulting equation of motion is precisely the Dirac equation.

\subsubsection{The classical limit and the Koopman--von Neumann perspective}

A central virtue of the ODM framework is that the classical limit is implemented in a
transparent and algebraically clean manner: one simply replaces the canonical commutation
relation with the commutative condition
\begin{equation}
\label{eq37}
[\hat{ x}^{j} , \hat{ p}_{k} ] = 0.
\end{equation}
In this case, the algebra must be extended to include auxiliary operators
$\hat{\theta}^k$ and $\hat{\lambda}_k$ satisfying
\begin{equation}
\label{eq38}
  {[}  \hat{x}^{j} ,  \hat{\lambda}_k {]} = -i \delta^{j}_{k} , \qquad
  {[}  \hat{p}_{j} ,  \hat{\theta}^k {]} = -i \delta^{k}_{j},
\end{equation}
as in the nonrelativistic Koopman--von Neumann (KvN) theory~\cite{Real-27,Real-28}.
However, the straightforward classical model obtained this way generates antiparticles
from a purely particle initial state, conflicting with classical physics.

The remedy is to modify the Ehrenfest relations by inserting the positive-energy projector
\begin{equation}
\label{eq39}
\mathcal{P}_{+} = \frac{1}{2} \left(
  \mathbf{1} + \frac{-\gamma^0\gamma^k c( \hat{p}_k - e A_k)  + mc^2 \gamma^0 }{ K(p)   }
\right),
\end{equation}
where $K(p) = \sqrt{ (mc^2)^2 +  c^2(p  - e A)^k\cdot(p - e A)^k }$ is the classical
kinetic energy. The resulting equation of motion,
\begin{equation}
\label{eq40}
 i \frac{\partial }{\partial t} \mathcal{W} =
 \frac{1}{2} \mathcal{P}_{+}  {[}  \gamma^0 \gamma^{\nu} , \hat{K}_{\nu} \mathcal{W}  {]}_{+} \mathcal{P}_{+},
\end{equation}
preserves the particle sector and agrees with classical Hamiltonian evolution.
This equation was originally derived by Spohn~\cite{Real-23}
from a semiclassical analysis, and is consistent with the
Bargmann--Michel--Telegdi equation for classical spin~\cite{Real-29}.
Within the ODM perspective, it is naturally identified as the relativistic
Koopman--von Neumann theory underlying the Dirac equation.

\subsubsection{Concluding comments on derivation via Operational Dynamical Modeling}

The ODM derivation clarifies the logical structure of the quantum--classical divide
in relativistic mechanics. The \emph{sole} algebraic distinction between the
quantum Dirac equation and its classical counterpart (Spohn's equation) is the
commutativity versus noncommutativity of position and momentum---the same
conclusion reached in the nonrelativistic setting~\cite{Real-21,Real-30}.
The classical limit additionally requires the explicit exclusion of negative-energy
(antiparticle) states. This approach avoids the lengthy algebraic manipulations
typical of earlier semiclassical treatments~\cite{Real-31,Real-32,Real-33,Real-34} and~\cite{Real-23}, which
provides a unified operational perspective linking the Dirac equation to
classical spinorial dynamics.

\subsection{Recent derivation from the Madelung equation}

A conceptually different route to the Dirac equation has recently been established through the hydrodynamic formulation of quantum mechanics introduced by Erwin Madelung~\cite{Real-36}. In~\cite{Real-35}, the equivalence between the Dirac equation in polar form and the Madelung equations was demonstrated. The deep connection between the Madelung formulation and classical hydrodynamics is discussed, e.g., in the monograph~\cite{Real-37}. It is worth recalling that Madelung originally derived his equations on the basis of the Schr\"odinger equation.

The Madelung system consists of real, non-linear partial differential equations of the form
\begin{equation}
\label{eq41}
m\dot{\vec{X}}=\vec{F}+\frac{\hbar^{2}}{2m}\nabla\frac{\triangle\sqrt{\rho}}{\sqrt{\rho}}, \quad \nabla \times \vec{X}=0, \quad \frac{\partial \rho}{\partial t}+\nabla \cdot (\rho \vec{X})=0,
\end{equation}
where $X=\frac{\partial}{\partial t}+\vec{X}$ is a real vector field, known as the drift (velocity) field, and $\dot{\vec{X}}$ denotes the material derivative of $X$ along itself:
\begin{equation}
\label{eq42}
\dot{\vec{X}}=\frac{\partial}{\partial t}+\frac{d \vec{X}}{dt}(t).
\end{equation}
Here $\rho$ is the probability density, $\vec{F}$ is the external force, and $m$ is the mass of the particle.

The equivalence of the Dirac and Madelung equations~(\ref{eq41}) was established in~\cite{Real-35}, with detailed proofs given both in the main text and in the appendices. The analysis covers the cases of two, three, and four dimensions. The path to this result proceeds through several intermediate stages: first the well-known relationship between the Madelung and Schr\"odinger equations, then the framework of relativistic hydrodynamics~\cite{Real-37,Real-38,Real-39,Real-40,Real-41,Real-42}, and finally the direct connection between the Madelung and Dirac equations originally explored by Yvon~\cite{Real-43}.

In conclusion, it is worth noting that the Madelung hydrodynamic approach has recently been undergoing a renaissance, furnishing new perspectives on Born-Oppenheimer molecular dynamics~\cite{Real-48}, spin-orbit coupling~\cite{Real-49}, and quantum-classical correspondence in nonadiabatic dynamics~\cite{Real-50}.

\section{Conclusions}

The centennial of quantum mechanics invites not only celebration but critical
reappraisal.  The historical record reconstructed above demonstrates that
the creation of relativistic quantum mechanics was not the work of a single
mind: Kramers' independent derivation~\cite{Real-14}, van der Waerden's
group-theoretical approach~\cite{Real-17,Real-18}, and Darwin's physical
elucidation~\cite{Real-10} each illuminate facets of the Dirac equation that its
originator's compressed presentation left implicit.  Kramers' seven-year
delay in publication, due in no small part to Pauli's skepticism, should
not obscure the depth of his contribution, and a faithful account of the
genesis of quantum theory must restore these figures to their proper place.

At the same time, the continuing appearance of genuinely new derivations
testifies to the equation's inexhaustible richness.  The Operational
Dynamical Modeling framework~\cite{Real-20,Real-21,Real-22} reveals that the \emph{sole} algebraic
distinction between quantum and classical relativistic dynamics is the
commutativity or noncommutativity of position and momentum, while the
Madelung hydrodynamic route~\cite{Real-35,Real-36} connects the Dirac equation to the
language of fluid mechanics.  Together with the 39 derivations catalogued
in~\cite{Real-7}, these results confirm that the full conceptual landscape of the
Dirac equation has yet to be mapped.

Looking ahead, the equation's relevance is undiminished.  Even its simplest
massless limit suffices to explain the physics of graphene nanoribbons and
topological materials, and its algebraic structure informs quantum computing. Moreover, the Dirac equation has enjoyed a renewed surge of
interest driven by the development of novel techniques---most notably the relativistic dynamical
inversion method---for constructing exact analytic solutions across a rich
landscape of problems, ranging from coherent control~\cite{Real-51} and electron vortex
beams in external electromagnetic fields~\cite{Real-52,Real-53} to
gravitational plane waves and near-horizon black-hole
dynamics~\cite{Real-54,Real-55,Real-56}.
At the same time, quantum theory confronts profound open questions---the
location of the quantum--classical boundary, the quantization of gravity, and
the nature of dark matter and dark energy---that will almost certainly require
extensions of the framework erected a century ago.  As Dirac himself urged in
his later writings~\cite{Real-44,Real-45,Real-46,Real-47}, the beauty of a physical theory may serve as a
guide to its correctness; a century of development has only strengthened that
conviction.

\section*{Acknowledgments}

V.S.\ is sincerely grateful to Valeriy Gusynin for a series of fruitful discussions concerning the formative period of quantum mechanics (1925--1928) and the contributions of Wolfgang Pauli, Hendrik Kramers and Vladimir Fock to the construction of the foundations of quantum theory.

D.I.B. was supported by DEVCOM Army Research Office (ARO) (grant W911NF-23-1-0288; program manager Dr.~James Joseph). The views and conclusions contained in this document are those of the authors and should not be interpreted as representing the official policies, either expressed or implied, of ARO or the U.S. Government. The U.S. Government is authorized to reproduce and distribute reprints for Government purposes notwithstanding any copyright notation herein.

\end{document}